\documentclass[showpacs,preprintnumbers,amsmath,amssymb]{revtex4}

\usepackage{graphicx}
\usepackage{dcolumn}
\usepackage{bm}

\usepackage{latexsym}
\usepackage{amsmath}
\usepackage{amssymb}
\usepackage{amsbsy}

\begin{document}

\title{Zero mode effect in the $1^{-+}$ four quark states}
\author{Z.F.Zhang$^1$ and H.Y.Jin$^{1,2}$}
\affiliation{ Zhejiang Institute
of Modern Physics,  Department of Physics, Zhejiang University, Zhejiang Province, People's Republic of China$^1$\\
CTP, Massachusetts Institute of Technology, Cambridge, MA 02139-4307$^2$}

\begin{abstract}
We calculate the masses of the $1^{-+}$ four quark states which decay
dominantly into
$\rho\pi$ and $\eta\pi$ respectively  by QCD sum
rules approach. We include the zero mode contribution and find it
plays an important role in the sum rules. We predict that the masses of
the states
$\eta\pi$ and   $\rho\pi$ both are 1.4-1.5 GeV. This is
close to the experimental candidates $\pi_1(1370)$ and $\pi_1(1440)$.
\end{abstract}


\maketitle

\section{INTRODUCTION}

$1^{-+}$ exotic mesons have been identified for several years.
In 1997, a group at BNL first claimed an
isopin vector meson $\pi_1(1370)$ with quantum number $1^{-+}$ in
the channel $\pi^- p\rightarrow \pi^- \eta p$\cite{BNL1}. This
state has been confirmed further by VES and CBar\cite{CBar}.
It could be identified as a hybrid at first sight.
However, all theoretical calculations show it seems not in the
case. The lattice QCD and the flux tube model predict the $1^{-+}$
mass is around 1.9GeV\cite{LATT}\cite{FLUX}, which is much heavier
than the $\pi_1(1370)$. QCD sum rule also predicts it should be
larger than 1.6GeV\cite{QCD}. Besides, the decays of the $1^{-+}$
hybrid have also been studied in the context of various
model\cite{FLUX}\cite{MODEL}, and also appear to be in
disagreement with the experimental data of the $\pi_1(1370)$. For
instance, the flux model predicts the $1^{-+}$ hybrid dominantly
decays into  $f_1\pi$ and  $b \pi$ and QCD sum rule's calculation
shows (although  differs from that of the flux model) $f_1\pi$ and $\rho\pi$
are the hybrid's dominant decay  channels. Most recently, Klempt gave a stronger
argument based on the SU(3) flavor symmetry to rule out the posibility
that the $\pi_1(1370)$ could be a hybrid\cite{KLEMPT}.  Because the  $\pi_1(1370)$
is seen in the channel $\eta\pi$ but not $\eta^\prime\pi$, the $\pi_1(1370)$
must be  a member of the SU(3) decuplet. Therefore, it could not be a $\bar
qqg$ hybrid.

An arugment is given by Close that the lowest $1^{-+}$ four quark state
should be in the combination $|0^-1^+\rangle$, such as $\pi f_1$ or $\pi
b_1$, because they are in S-wave\cite{close}. However , both of them are heavier than
the $\pi_1(1370)$ and consequently cannot be seen in the $\pi_1(1370)$ decay.
Although  $\eta\pi$ and $\rho\pi$ are  in P-wave,  the combinations
$\eta\pi$, $\rho\pi$, $\pi f_1$ and $\pi b_1$ belong to the same order of orbital
excitation for a four quark system. Their masses should not be quite
different.  $\eta\pi$ is already seen in the $\pi_1(1370)$
decay. But why not for $\rho\pi$ ? Actually, when $\pi^0\eta$ does not
appear in BNL's reaction  $\pi^-p\rightarrow \pi^-\eta p$, some authors
conclude that  the $1^{-+}$ $\pi_1(1370)$
might not be exist\cite{AR}. Klempt's answer is that in the t-channel there is no
$\rho$ exchange. That means the $\pi_1(1370)$ couples $\rho\pi$ very weakly.
Then how about the $\pi_1(1440)$?  Which is another $1^{-+}$ state seen in the $\rho\pi$
channel and
a little bit heavier than the $\pi_1(1370)$\cite{WF}. Is the $\pi_1(1440)$ the same
 as the $\pi_1(1370)$ ?
If yes, it contradicts to Klempt's statement. If no, how to explain the two
states with the same quantum number and only 70$MeV$ mass interval ?

In order to investigate this problem, we calculate the masses of the
$1^{-+}$ four states which decay dominantly into $\rho\pi$ and $\eta\pi$
respectively
from QCD sum rule approach. We find the direct instanton effect plays an
important
role. The direct instanton effect is very large in the sum rules of  the
state $\eta\pi$ while it is
proportional to the light quark mass square for the case $\rho\pi$. This
might hint the different
structure of states $\rho\pi$ and $\eta\pi$. The predicted masses  of
$\rho\pi$  and $\eta\pi$
are  both 1.4-1.5 GeV, which is close to the
experimental candidates $\pi_1(1370)$ and $\pi_1(1440)$.


\section{SUM RULES FOR $1^{-+}$ FOUR QUARK STATES}

The main task for the $1^{-+}$ mass prediction in the QCD sum rule
approach is to calculate the current-current correlator
\begin{equation}
\begin{split}
\Pi_{\mu\nu} (q^2) &= i \int d^4x e^{i q x} \langle 0|T j_\mu (x)
j_\nu^\dagger (0) | 0 \rangle \\
&=\left( \frac{ q^\mu q^\nu} {q^2} - g^{\mu \nu} \right) \Pi_v
(s^2)
+ \frac{ q^\mu q^\nu} {q^2} \Pi_s(s^2),
\end{split}
\end{equation}

Where the interpolated current $j_\mu (x)$ has the quantum number
$1^{-+}$. In this paper, we just forcus on $\pi_1(1370)$ and $\pi_1(1440)$.
These two states
are only seen in the channel $\pi\eta$ and $\pi\rho$  respectively.
They look like molecule states. We need to  construct the  $1^{-+}$
four quark currents with such property, for instance,
\begin{equation}
j_ {1\mu} = \bar q(x) \gamma _ 5 \sigma q(x) \bar q(x) \gamma _ 5
\gamma _ \mu q(x)
\end{equation}
for the $\eta\pi$ state (where  $\sigma$ is the isospin matrix and
$q=2^{-1/4} ( u , d )^T$), and
\begin{equation}
j_{2\mu\nu} = \epsilon_{\mu\nu\rho\sigma}(\bar u\gamma _ 5\gamma^\rho
d\bar d \gamma ^ \sigma u-\bar d\gamma _ 5\gamma^\rho
u\bar u \gamma ^ \sigma d)
\end{equation}
for the $\rho\pi$ state. We cannot find a dimension six $1^{-+}$ current for
the $\rho\pi$ state. But $j_{2\mu\nu}$ indeed annihilates a  $1^{-+}$ $\rho\pi$ state.

These currents do not exactly represent the molecule states.
For instance, $j_{1\mu}$ can both easily decay into
$\pi\eta(\eta^\prime)$ and $\pi f_1$ if its mass permits. We cannot avoid
such property, but other
channles, such as $b_1\pi$ and $\rho\pi$,  is indeed  suppressed in
$j_{1\mu}$'s  decay.

Use the standard operator-product expansion (OPE)
method\cite{novikov}, we get (up to irrelevant polynomials in $
q^2 $)
\begin{gather}
\label{eq:nzm}
{\rm Im} \Pi (s )_{1v}^{(OPE)} =  \frac{11} {1179648}
\frac{s^4}{\pi^5}
+ \frac{\langle \alpha_s G^2\rangle } {8192} \frac{s^2}{ \pi^4}
- \frac{\langle \bar qq \rangle^2}{128} \frac{s}{\pi},\\
{\rm Im} \Pi_{2v}^{(OPE)} (s) = \frac{1}{30720 \pi^5} s^4 -\frac{\langle \alpha_s G^2 \rangle}
{768 \pi^4} s^2 +
\frac{5}{324} \frac{s}{\pi^2} \alpha_s(s) \langle \bar q q \rangle^2
\left[ 1+6 \gamma_E +12 \ln (s/\mu^2)\right].
\end{gather}
where $\gamma_E$ is Euler's constant and we will use
$$\alpha_s(q^2)= \frac{4 \pi}{9 \ln(q^2/\mu^2)},$$
we also have ignored the two-quark condensate since it always
accompanys with the mass of light quark, thus it is less
importmant compare with these gluon condensate. In the correlator of
the current $j_{2\mu\nu}$, the leading order four quark condensate and
the three gluon condensate vanish.  The next leading order  four quark condensate is very important.
This situation is similar to that of the meson $\rho$.

In order to preform QCD Sum Rules for the $1^{-+}$ four quark
states, we also should know something about the meson spectral
density. Usually the spectral density $ \rho_v(s) = {\rm Im} \Pi_v
(s)$ is defined via the standard dispersion relation
\begin{equation}
\label{eq:dis} \Pi_v (q^2) = \frac{ (q^2)^n} {\pi} \int^\infty_0
ds \frac{ \rho_v(s)} {s^n(s-q^2)} + \sum\limits_{k=0}^{n-1} a_k
(q^2)^k,
\end{equation}
where the $a_k$ are appropriate subtraction constants to render
Eq.\ref{eq:dis} finite.

After Borel transforming the spectral density, we get the Sum
Rule:
\begin{equation}
R_i (M_B^2) =\frac{1}{\pi} \int^{s_0}_0  e^{-s /M_B^2} s^i \rho_v (s) ds,
\end{equation}
where $i=1,2$, and the quantity $R_k$ represents the QCD
prediction, and the threshold $s_0$ separates the contribution
from higher excited states and the QCD continuum.

In the single narrow resonance scenario, the lowest-lying
resonance mass can be obtained from ratios
\begin{equation}
\label{eq:sumrule} m_v^2 = \frac{ \int^{s_0}_0 e^{-s/M_B^2} s^{2}
\rho_v (s) ds}
{\int^{s_0}_0 e^{-s /M_B^2} s \rho_v (s) ds}.
\end{equation}

Thus we can use Eq.\ref{eq:sumrule} to predict the mass of the
$1^{-+}$ four quark state.
\begin{figure}[htbp]
\centering
\includegraphics{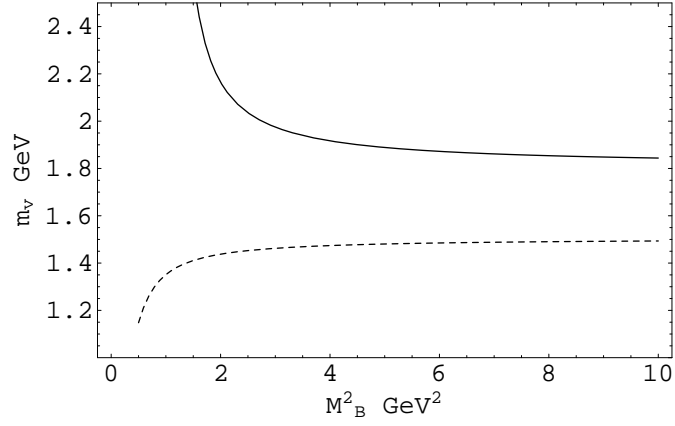}
\caption{\label{fig:sr1} Sum Rules for the $\eta\pi$ and $\rho\pi$ states, where
$s_0=3GeV$. ( The solid lines correspond to the Sum Rules
for $\eta\pi$ while the dashed lines the Sum Rules for the $\rho\pi$)}
\end{figure}

The various QCD parameters that will be used in the
phenomenological analysis of Eq.\ref{eq:sumrule} are
\begin{gather}
\langle \alpha_s G^2 \rangle = 0.08 GeV^4,\\
\langle \bar q q\rangle = - ( 0.24 GeV)^3,\\
\mu=0.2 GeV.
\end{gather}

In Fig.\ref{fig:sr1},  we show the mass of $1^{-+}$ four quark
state refered to the Borel parameter $M_B$.

The sum rules of the  $\eta\pi$ state  are  not stable(at small Bore parameter the spectrum
density is negative) and the predicted
mass of the $\eta\pi$
is heavier than that of the  $\rho\pi$. This is contradictory to the
experiments about the
$\pi_1(1370)$ and the $\pi_1(1440)$. This problem arises because  we have not
taken the direct
instanton effect into account yet. In history, QCD sum rules based on OPE
gave a good description of
vector mesons, such as  $\rho$, $\omega$, $J/\psi$ and so on, but failed
in scalar or pseudoscalar mesons. It was found later that the direct
instanton effect is large in the scalar(pseudoscalar) channel
but little in the vector channel. However, this statement is only  valid in
the quark anti-quark system. For a four quark system, we can see the direct
instanton effect could be also very important in the vector channel.

The calculation of the direct instanton effect is via the so-called  zero
mode($\psi_0$), which
is a classical solution of the Dirac equation (given by $^\prime$t Hooft)
in the background field of instantons:$D\!\!\!\!/ \psi_0=0$.

Expanding the quark propagator in the background field of
instanton for small quark mass, we get \cite{shuryak2}:
\begin{equation}
\label{eq:full} S(x,y) = \frac{ \psi_0 ( x ) \psi_0^\dagger ( y )}
{i m }
+S^{nzm} ( x , y) + m \Delta ( x , y ) + \cdots,
\end{equation}
where $\Delta ( x , y )$ is the propagator of a scalar quark.

The first term of the expansion is known as the zero mode part of
the propagator in the instanton field, which reads
\begin{equation}
S^{zm} (x,y;z) = \frac{(x\!\!\!/ -z\!\!\!/ ) \gamma_\mu \gamma_\nu
(y\!\!\!/ - z\!\!\!/ )}{8 m^* i} \left[ \tau^-_\mu \tau^+_\nu \frac{1-
\gamma_5}{2} \right] \phi(x-z) \phi(y-z),
\end{equation}
where $$\phi(x)=\frac{\rho}{\pi}
\frac{1}{|x|(x^2+\rho^2)^{3/2}},$$ and $\tau^{\pm}_\mu = (
\tau,\mp i)$.

Now let us consider the case of the current $j_{1\mu}$.
The leading contributions of non-zero mode have already obtained
in Eq.\ref{eq:nzm}. From Eq.\ref{eq:full} we know that the zero
mode part is the dominating part of the full propagator since for
light quark, $m^*$ is very small. So Eq.\ref{eq:nzm} is not a
complete correlation function. We need several new diagrams which
include zero mode contribution.
\begin{figure}[htbp]
\centering
\includegraphics[height=3cm]{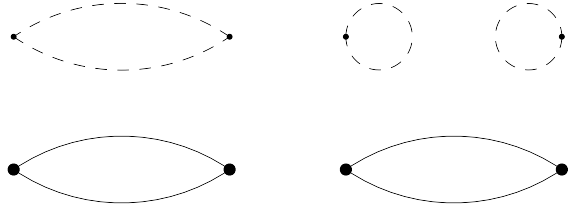}
\caption{\label{fig:zeromode}Zero mode contribution figure (I).
(The bigger blob denotes the pseudo-vector vertex
$\gamma_5\gamma_\mu$, while the smaller one is the pseudo-scalar
vertex $\gamma_5$. The dashed lines mean zero mode propagator. The
coordinates of the left two vertices are zero while the right ones
x.) }
\end{figure}

First we must include Fig.\ref{fig:zeromode}, in which the
pseudo-scalar loop receives zero mode contributions. In the single
instanton approximation, the result  is
\begin{equation}
\Pi^{SIA}_{ps} (x) =  - \frac{6n\rho^4}{{m^*}^2\pi^2} \int_0^1 dy
\frac{y^2(1-y)^2} {\left[ x^2y(1-y)+\rho^2\right]^4},
\end{equation}
where the effective mass is $m^*=\pi \rho (2 n /3)^{1/2}$
according to the mean field estimate\cite{shuryak}. This function
is well defined. It vanishes as $x$ goes to infinity, meanwhile it
is finite when $x$ goes to zero. In order to simplify our
calculation, we expand this function about the point $x=0$ and
only preserve the leading order, that is, we only preserve the
constant term. Combining with the pseudo-vector loop contribution,
we finally get the zero mode contribution for
Fig.\ref{fig:zeromode}.

The pseudo-vector loop does not received contribution of zero
modes, though it does received a contribution from the
interference between the zero mode part and the lead mass
correction, we ignore it either since the effect is not very
importmant.
\begin{figure}[htbp]
\centering
\includegraphics[height=5cm]{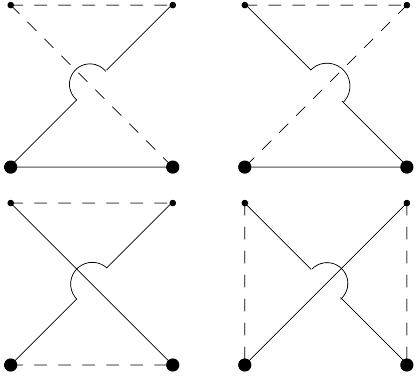}
\caption{\label{fig:zeromode2}Zero mode contribution figure (II).}
\end{figure}

Besides Fig.\ref{fig:zeromode}, there are still several diagrams,
showing in Fig.\ref{fig:zeromode2}, which also receive zero mode
contribution. The contribution of Fig.\ref{fig:zeromode2} must be
doubled since the arrows of the fermion lines have two direction.

After combining all contribution of Fig.\ref{fig:zeromode} and
\ref{fig:zeromode2}, we finally get:
\begin{equation}
\label{eq:zeromode}
{\rm Im} \Pi _{1v}^{(zm)} (s) = \frac { 21} {320} \frac{s} {
\pi^5\rho^6},
\end{equation}
where all zero mode contributions preserve to leading order.

\begin{figure}[htbp]
\centering
\includegraphics{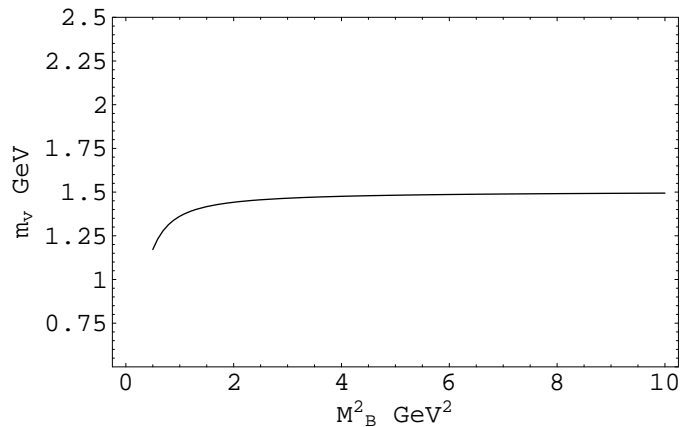}
\caption{\label{fig:sr2} Sum Rules for the $\eta\pi$ includes zeromode contributions, where
$s_0=3GeV$.}
\end{figure}

Combinging Eq.\ref{eq:nzm} and Eq.\ref{eq:zeromode} and using the
new  parameters $\rho= 1/ 0.6 GeV$ we finally get Fig.\ref{fig:sr2}
, which show the mass of the resonance is  1.4-1.5 GeV.

Because the zero mode flips helicity of the quark, if all vertices of the current
are vector type, the direct instanton effect via the zero mode is proportional to the light quark
mass (or higher power), because it needs the quark mass to flip helicity. Therefore
for the second current, the direct instanton contribution
from the zero mode is proportional to the light quark mass square. We neglect it.

\section{DISCUSSION AND SUMMARY}
After taking account of the direct instanton contribution, we find
a dramatic  result that the sum rules for the $\eta\pi$ become quite stable.
From the Fig.\ref{fig:sr2}, we read the mass of the $\eta\pi$ is around 1.4-1.5
GeV. Because the correlator of the current $j_{2\mu\nu}$  does not get a big
direct instanton contribution, the mass of $\rho\pi$ is still 1.4-1.5GeV. The mass of
the $\eta\pi$ is slightly lighter than that of the $\rho\pi$.
This is quite consistent with the experimental data of the $\pi_1(1370)$ and
$\pi_1(1440)$. Moreover, the different impact of the direct instanton
contribution
on the correlators of $j_{1\mu}$ and $j_{2\mu\nu}$ probably hints the
different
sturcture of states $\eta\pi$ and $\rho\pi$. One might think the difference
of the direct instanton contribution is compensated by the difference of
the four quark condensate, because they have the same dimension. But we also
note that
the sign of the four dimension gluonic condensate of these two correlators
is  opposite. All of these hints the mixing between the $\eta\pi$ and
$\rho\pi$
might be small. This result should not be quite surprising. In our opinion, because 
a four quark system has much more degree of freedom than a two quark system, the spectrum of 
the four quark system should be more crowded. This has  already been comfirmed in the Ref.\cite{dorokhov}, where 
the authors find there are several four quark states with the same quantum number and some of them are almost degenerate. Besides,
we need to mention that we only consider the two flavor
case. If we
include $s$ quark, the mass prediction could be slightly different. For
instance,
it is nature that the state $\eta^\prime\pi$ might be 200-300MeV heavier.
However, such calculation is more complicated. Besides, our calculation
assumes the states have
the molecule structure. If states have the different structure, such as the
diquark
structure, the prediction probably is different. We will discuss such cases
in another
paper.

\begin{acknowledgments}
The work was supported by NSFC and ZJNSF. H.Y. Jin thanks CTP's the warm hospitality.
\end{acknowledgments}

\end{document}